\documentclass[twocolumn,prl,showpacs,nofootinbib,superscriptaddress]{revtex4}
\usepackage{latexsym}
\usepackage{dcolumn}
\usepackage{epsfig}
\RequirePackage{graphicx}

\newcommand{\ignore}[1]{}
\newcommand{\ba}{\begin{eqnarray}}

\newcommand{\ea}{\end{eqnarray}}
\newcommand{\be}{\begin{equation}}
\newcommand{\ee}{\end{equation}}

\newcommand{\eq}[1]{Eq.\,(\ref{#1})}

\newcommand{\lessabout}{\raisebox{-.6ex}{\ $\stackrel{<}{\sim }$\ }}
\newcommand{\greaterabout}{\raisebox{-.6ex}{\ $\stackrel{>}{\sim }$\ }}

\def\bea{\begin{eqnarray}} 
\def\eea{\end{eqnarray}} 
\def\rd{{\mathrm d}} 
 
\def\intd4x{\int{\rd}^4x}

\def\m32{{m_{3/2}}}

\newcommand{\la}{\raisebox{-.8ex}{\,$\stackrel{\textstyle <}{\sim}$}\,} 
\newcommand{\ga}{\raisebox{-.8ex}{\,$\stackrel{\textstyle >}{\sim}$}\,}

\begin{document}

\preprint{ANL-HEP-PR-07-55}

\title{Ultra-high energy neutrino scattering}

\author{Edmond~L.~Berger}
\affiliation{High Energy Physics Division,
Argonne National Laboratory,
Argonne, Illinois 60439}
\author{Martin~M.~Block}
\affiliation{Department of Physics and Astronomy, Northwestern University, 
Evanston, IL 60208}
\author{Douglas W. McKay} 
\affiliation{Department of Physics and Astronomy, University of Kansas, Lawrence, KS 66045}
\author{Chung-I Tan}
\affiliation{Physics Department, Brown University, Providence, RI 02912} 
\date{\today}

\begin{abstract}
 
 Estimates are made of ultra-high energy neutrino cross sections based on an extrapolation to 
 very small Bjorken $x$ of the logarithmic Froissart dependence in $x$ shown previously to provide 
 an excellent fit to the measured proton structure function $F_2^p(x,Q^2)$ over a broad range of 
 the virtuality $Q^2$.  Expressions are obtained for both the neutral current and the charged 
 current cross sections.  Comparison with an extrapolation based on perturbative QCD shows good 
 agreement for energies where both fit data, but our rates are as much as a factor of 10 smaller 
 for neutrino energies above $10^9$~GeV, with important implications for experiments  searching  for 
 extra-galactic neutrinos.   

\end{abstract}


\maketitle
{\em Introduction:} The experimental effort to detect extra-galactic, ultra-high-energy (UHE) neutrinos has grown 
rapidly in the past decade. Optical~\cite{optical} and radio~\cite{radio} telescopes and cosmic ray air shower 
arrays~\cite{hires-auger} are now searching for evidence of point and diffuse neutrino sources up to and beyond EeV 
energies. Proposals have been made and others are in preparation~\cite{salsa-aura} for new telescopes or expansions 
of ones currently deployed, and ambitious satellite-born telescopes have been proposed~\cite{euso}.  The highest 
energies proposed reach beyond $10^{12}$ GeV.

Critical to all of this effort are accurate estimates of event rates, based on the extrapolation of measured neutrino 
deep-inelastic scattering (DIS) cross sections to energies 
{\it far} beyond currently available data~\cite{QCDnu,gandhi,mckay}.  The estimates are only as reliable as the 
extrapolations, and determination of fluxes and extraction of signals 
of new physics at UHE depend on them.  Most existing extrapolations are done within the framework of perturbative 
quantum chromodynamics (pQCD), and they involve extending fitted parton distribution functions (PDFs) into 
domains in Bjorken $x$ much below those now accessible experimentally, and into domains in which 
linear pQCD evolution~\cite{dglap} is of questionable applicability.  Other physical phenomena are expected 
to alter the $x$ dependence in this very small $x$ region~\cite{glr-capella-soyez}, 
although a complete analytic solution does not yet exist.     

New, alternative methods of extrapolation in $x$ are of significant interest, both theoretically and for phenomenological 
applications. Imposition of the Froissart~\cite{froissart} unitarity and analyticity constraints on inclusive deep-inelastic 
cross sections~\cite{bbt} leads to the expectation that the $x$ dependence of the proton structure function 
$F_2^p(x,Q^2)$ should grow no more rapidly at very small $x$ than $\ln^2(1/x)$.  This relatively slow growth may be 
contrasted 
with the more rapid inverse power dependence characteristic of PDFs.  Excellent fits to data were obtained~\cite{bbt} for 
$x < 0.1$ with an assumed logarithmic expansion, for a wide range of virtuality $Q^2$.  We explore in this Letter the 
consequences of the Froissart logarithmic form for UHE neutrino phenomena, computing both
neutral and 
charged current cross sections. In doing so, rather than working with parton distribution functions for the decomposition 
into quark and antiquark contributions, we devise and test a procedure based {\em directly} on experimental $F_2^p$ data.  
We obtain excellent 
agreement with extrapolations based on the CTEQ4-DIS parton densities in the neutrino energy range less than $10^8$~GeV.  
However, we predict an important departure for larger energies, with our neutrino cross sections being about a decade 
smaller at the highest energies.  At the very least, our results suggest that estimates that fall between ours and those 
obtained from PDF extrapolations be used for guidance in the consideration of new experiments.    

{\em  Neutrino-isoscalar nucleon cross sections:}
In the standard parton model the inclusive differential cross section for the charged current (CC) reaction 
$\nu_{\ell}+ N\rightarrow \ell^- + X$
on an isoscalar nucleon $N=(n+p)/2$ and the neutral current (NC) 
cross section $\nu_\ell+N\rightarrow\nu_\ell+X$, where in both cases, $\ell=e,\mu,\tau$, 
is
\ba
\frac{d^2\sigma}{dxdy}(E_\nu)&=&
\frac{
2G_F^2mE_\nu}{\pi}\left(\frac{M_V^2}{Q^2+M_V^2}\right)^2\times \nonumber\\
&&\left[xq_i(x,Q^2)+x\bar q_i(x,Q^2)(1-y)^2\right],\label{CC}
\ea
where $-Q^2$ is the invariant squared momentum transfer between the incoming neutrino and the outgoing muon, 
$m$ is the proton mass, and $G_F$ is the Fermi 
coupling constant.  The intermediate vector boson mass, $M_V$,  is $M_W=80.4$ GeV for CC and $M_Z=91.2$ 
GeV for NC.  Symbols $q_i$ and $\bar q_i$, $i=$CC, NC, are linear combinations of quark and antiquark PDFs.  
The Bjorken scaling variables, where $\nu=E_\nu-E_\ell$ is the energy loss in the laboratory frame, are given by
\be 
x\equiv\frac{Q^2}{2m\nu}, \quad y\equiv\frac{\nu}{E_\nu}, \quad 0\le x,y\le 1.
\ee

{\em  Charged current cross section}: 
With valence and sea quark distributions denoted by subscripts $v$ and $s$, respectively,  
the relevant PDFs in \eq{CC} are 
\ba
q_{\rm CC}(x, Q^2)&=& \frac{u_v(x, Q^2)+d_v(x, Q^2)}{2}\nonumber\\
&&+\frac{u_s(x, Q^2)+d_s(x, Q^2)}{2}\nonumber\\
&&+s_s(x, Q^2)+b_s(x, Q^2)\label{q}
\ea
and
\ba
\bar q_{\rm CC}(x, Q^2)&=& \frac{u_s(x, Q^2)+d_s(x, Q^2)}{2}\nonumber\\
&&+c_s(x, Q^2)+t_s(x, Q^2)\label{qbar}, 
\ea 
where $u$, $d$, $c$, $s$, $t$, and $b$ represent the contributions from the 
up, down, charm, strange, top, and bottom flavors.  

{\em Neutral current cross section}: The relevant PDFs in \eq{CC} involve 
chiral couplings
$
L_u=1-\frac{4}{3}\sin^2\theta_W,
L_d=-1+\frac{2}{3}\sin^2\theta_W,
R_u=-\frac{4}{3}\sin^2\theta_W,
R_d=\frac{2}{3}\sin^2\theta_W, 
$
where $\sin^2\theta_W=0.226$ is the weak mixing parameter. For details, see Ref. \cite{gandhi}.

{\em Kinematics:}  
Replacing $Q^2$ in \eq{CC} by  
$
Q^2=2mE_\nu x y\label{Qsq},
$
we obtain an expression in terms of $E_\nu, x$ and $y$.  We choose to integrate first over $y$. To avoid singularities in the integration, 
we  introduce  $Q^2_{\rm min}=0.01$ GeV$^2$, such that $Q^2=2mE_\nu xy \ge Q^2_{\rm min}$.  This defines $x_{\rm min}$, 
the $x$-integration minimum, as
$
x_{\rm min}\equiv Q^2_{\rm min}/(2mE_\nu). \label{xmin}
$
Thus,  for $x_{\rm min}\le x\le 1$,  our integration limits for $y$ are $y_{\rm min}=x_{\rm min}/x \le y\le 1$. 

The vector boson  propagator, $\left(M_V^2/(Q^2+M_V^2)\right)^2$, 
essentially fixes an ``effective'' $x$   at $x_{\rm eff}\sim M_V^2/(2mE_\nu)$.  For 
$E_\nu=10^{12}$ GeV, this means we must explore quark distributions having $x_{\rm 
eff}\sim 5\times 10^{-9}$, at $Q^2\sim M_V^2\sim 10,000$ GeV$^2$, both of which involve 
{\em enormous} extrapolations from currently available  structure function data. 
At these energies, the propagator also serves to make the calculation insensitive to the choice of $Q^2_{\rm min}$.  

{\em  Analytic expression for the structure function}:  
In prior work~\cite{bbt}, it was shown that an excellent fit to the DIS structure 
function for $x\le x_P$, is given by
\ba
F_2^p(x,Q^2)&=&(1-x)\left( \frac{F_P}{1-x_P}+A(Q^2)\ln\left[\frac{x_P}{x}\frac{1-x}{1-x_P}\right]\right.\nonumber\\
&&\left.+B(Q^2)\ln^2\left[\frac{x_P}{x}\frac{1-x}{1-x_P}\right]\right),\label{Fp}
\ea
where 
\ba
A(Q^2)&=& a_0+a_1\ln Q^2+a_2\ln^2 Q^2,\nonumber\\
B(Q^2)&=& b_0+b_1\ln Q^2+b_2\ln^2 Q^2. \label{A&B}
\ea
The fitted numerical values of $a_j$ and $b_k$ and their uncertainties may be found 
in Ref.~\cite{bbt}; 
$F_P=0.41$, and $x_P=0.09$.

The bulk of the neutrino cross section comes from exceedingly small $x$. For large $x$, where  $x_P\le x\le 1$, it 
suffices to approximate the proton structure function by 
\be
F_2^p(x,Q^2)=\frac{F_P}{x_P^{\alpha(Q^2)}(1-x_P)^3}x^{\alpha(Q^2)}(1-x)^3 ,\label{Fhi}
\ee
where the exponent $\alpha(Q^2)$ is chosen so that the first derivatives of \eq{Fp} and \eq{Fhi} are equal at 
$x=x_P$. This choice satisfies the spectator valence quark counting rule~\cite{countingrule} 
$F^p_2(x)\rightarrow 0$ as $(1-x)^3$ as $x\rightarrow 1$.  
Numerical analysis shows that this choice  has the important consequence that the integral of the proton structure 
function over $x$ is nearly constant over an enormous $Q^2$ range,  i.e., 
\be
\int_0^1F_2^p(x,Q^2)\,dx \approx 0.16, \quad 0.1\le Q^2 \la 10^5{\rm \ GeV}^2 .\label{f2int}
\ee
The constant 0.16 is compatible with results that show that quarks carry $\sim  50\%$ of the 
momentum in a proton. 

The description of $F_2^p(x,Q^2)$ by Eqs. (5) - (8) yields a high quality fit to the HERA inclusive 
deep-inelastic data for all x and $Q^2$.

{\em ``Wee parton'' picture:}  
We obtain the quark distribution functions in \eq{CC} from a wee parton model for very small Bjorken $x$,  
having the following features:
\begin{itemize}
\item there are essentially {\em only} sea quarks at small enough $x$, with {\em negligible} valence quark contributions 
(for earlier use, see  Ref. \cite{mckay}), i.e., we set\\   
$
u_v(x,Q^2)=d_v(x,Q^2)=0.
\label{valence}
$
\item  all sea quarks give equal contribution 
(i.e., equipartition), 
\\
$
U(x,Q^2)=
u_s(x,Q^2)=\bar u_s(x,Q^2)=d_s(x,Q^2)
=\bar d_s(x,Q^2)
=s_s(x,Q^2)=\bar s_s(x,Q^2)
=c_s(x,Q^2)=\bar c_s(x,Q^2). \label{sea}
$
\end{itemize}

If only two families contribute ($u$, $d$, $c$, and $s$),
\ba
\!\!F_2^p(x,Q^2)\!=\!\!\sum_i e_i^2 x[q_i(x,Q^2)\!+\!\bar q_i(x, Q^2)],  i=1,\ldots 4 \label{quarkcontent}, 
\ea 
or, alternatively, 
\be
xU(x,Q^2)=\frac{9}{20}F_2^p(x,Q^2)\label{commonxU},
\ee
for $x < x_{\rm max}$, where $x_{\rm max}\sim 10^{-3}-10^{-4}$.
If we had used {\em only one family} of quarks---$u,d$---or 
{\em three families}---$u,d,c,s,t,b$---instead two families---$u,d,c,s$---
we would also find that $xq(x,Q^2)=x\bar q(x, Q^2)=\frac{9}{10} F_2^p(x,Q^2)$, so 
that \eq{CC} for charged currents is independent of the number of families. A 
similar result is true for the neutral current cross section. 
Employing this picture, we find that accurate knowledge of $F_2^p(x,Q^2)$ at small 
$x$ and large $Q^2$ provides the ingredients necessary to calculate the charged and neutral 
current neutrino cross sections.  The fitted form of \eq{Fp} is sufficiently 
accurate to furnish us with  quark distribution functions having the needed precision.  Using the 
full squared error matrix for the structure function determination~\cite{bbt}, 
we find that $F_2^p(x=10^{-8},Q^2=6400\ {\rm GeV}^2)=24.84\pm0.17$, a fractional {\em statistical} 
accuracy of only $\sim 0.7\%$.  This very small uncertainty due to {\em parameter errors}  assumes, of course, 
the validity of our $\ln^2(1/x)$ model  at very small $x$.  

{\em Charged current cross section evaluation:}  
For our model,
$xq_{\rm CC}(x, Q^2)=x\bar q_{\rm CC}(x, Q^2)=2xU(x,Q^2)$. 
Thus
$
xU(x,Q^2)=\frac{9}{20}F_2^p(x,Q^2)
\label{xU}
$
and 
 \eq{CC} simplifies to 
\ba
\frac{d^2\sigma_{\rm CC}}{dxdy}(E_\nu)&=&
\frac{
2G_F^2mE_\nu}{\pi}\left(\frac{M_W^2}{Q^2+M_W^2}\right)^2\times\nonumber\\
&&\left[\frac{9}{10}F_2^p(x,Q^2)\right](2-2y+y^2),\label{CC1}
\ea
with $F_2^p(x,Q^2)$ given by \eq{Fp} for $0\le x \le x_P$ and \eq{Fhi} for  $x_P<x\le 1$.

\begin{figure}
\begin{center}
\mbox{\epsfig{file=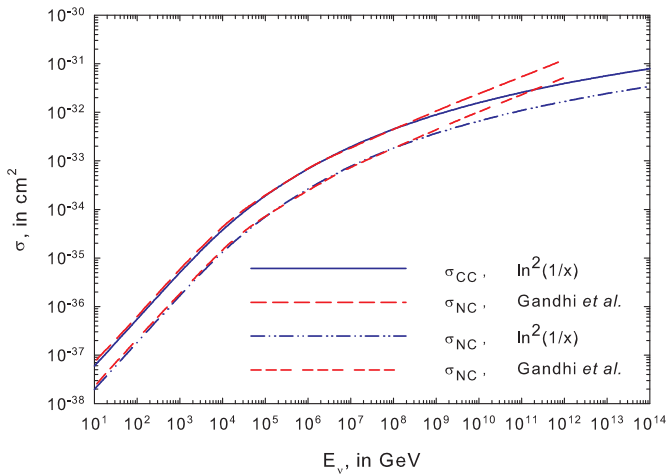,width=3.5in%
,bbllx=0pt,bblly=0pt,bburx=435pt,bbury=310pt,clip=%
}}
\end{center}
\caption[]{Charged (CC) and neutral (NC) current neutrino cross sections in cm$^2$ vs $E_\nu$, the neutrino 
energy in GeV. The solid and dash-dot-dot curves are  our CC and NC cross sections,  respectively, 
for $10\le E_\nu\le10^{14}$ GeV, based on a proton structure function that varies as $\ln^2(1/x)$ for small $x$. 
The long dash curve and the dash-dash-dash curve are the Gandhi {\em et al.} \cite{gandhi}  CC and NC cross 
sections , respectively,  
for $10\le E_\nu\le 10^{12}$ GeV, based on the CTEQ4-DIS quark distributions.
\label{fig:CC}}
\end{figure}

Results of a direct double integration of \eq{CC1}, 
with $Q^2_{\rm min}=0.01$ GeV$^2$, for the neutrino energy range $10\le E_\nu\le 10^{14}$ GeV, 
are given in Table \ref{crosssections} and shown in Fig. \ref{fig:CC} as the solid curve. Also 
shown, for comparison, are the results of Gandhi {\em et al.} \cite{gandhi} for the CC 
cross section 
with the quark distributions 
from CTEQ4-DIS\cite{CTEQ4}. 
The Gandhi {\em et al.} curve---the long dash curve---covers the energy range 
$10\le E_\nu\le 10^{12}$ GeV. The agreement up to neutrino energies 
$\lessabout 10^8$ GeV is striking.

{\em Neutral current cross section evaluation:}  
For our model, the NC quark distributions in \eq{CC} are
\ba
xq_{\rm NC}(x, Q^2)&=&x\bar q_{\rm NC}(x, Q^2)= 2xU(x,Q^2)\times\nonumber\\
&&\qquad\qquad(L_u^2+L_d^2+R_u^2+R_d^2)\nonumber\\
&=&4(1-2\sin^2\theta_w+\frac{20}{9}\sin^4\theta_W)xU(x,Q^2)\nonumber\\
&=&2.65xU(x,Q^2)
=1.19F_2^p(x,Q^2),\label{qNC}
\ea
where \eq{commonxU} is used in the last step.  The neutral current cross section simplifies considerably. For 
direct comparison with the charged current cross section of \eq{CC1}, it can be rewritten as
\ba
\frac{d^2\sigma_{\rm NC}}{dxdy}(E_\nu)&=&
\frac{
2G_F^2mE_\nu}{\pi}\left(\frac{M_Z^2}{Q^2+M_Z^2}\right)^2\times\nonumber\\
&&\left[0.298F_2^p(x,Q^2)\right](2-2y+y^2).\label{NC1}
\ea
To the extent that the $Z$ propagator is somewhat less restrictive as a cutoff than the $W$ propagator, 
comparison of \eq{NC1} and \eq{CC1} shows that the {\em ratio} of the NC cross section to the CC cross section 
is $\ga0.298/0.9= 0.33$, independent of energy.  Numerical evaluation gives $~0.40$ at $E_\nu=10^{7}$ GeV, slightly 
higher because of the $Z$ propagator. Our NC cross section for isoscalar nucleons is given in Table \ref{crosssections} 
and shown in Fig. \ref{fig:CC} as the dash-dot-dot curve, plotted in the energy interval $10\le E_\nu\le 10^{14}$ GeV.  
The Gandhi {\em et al.}~\cite{gandhi} NC cross section, for $10\le E_\nu\le 10^{12}$ GeV, is the dash-dash-dash curve.  
Again, the agreement is excellent up to $E_\nu\sim 10^8$ GeV.

\begin{table}[h] 
\begin{center}
\def\arraystretch{1.2}            
     \caption{\label{crosssections}\protect\small Neutrino CC and NC total cross sections, with neutrino energy $E_\nu$ 
energy in GeV and cross sections in cm$^2$.}
\begin{tabular}[b]{|l|c|c||l|c|c||}     
\hline
$E_\nu$&$\sigma_{\rm CC}$&$\sigma_{\rm NC}$&$E_\nu$&$\sigma_{\rm CC}$&$\sigma_{\rm NC}$\\
\hline
      $10^{1}$&$ 5.93\ 10^{-38}$&$1.96\ 10^{-38}$&$10^8$&$4.49\ 10^{-33}$&$1.83\ 10^{-33} $\\ 
      $10^2$&$ 5.51\ 10^{-37}$&$1.82\ 10^{-37}$&$10^9$&$8.90\ 10^{-33}$&$3.70\ 10^{-33}$\\
      $10^3$&$5.01\ 10^{-36}$&$1.6710^{-36}$ &$10^{10}$&$1.58\ 10^{-32}$&$6.63\ 10^{-33}$ \\
	$10^4$ &$3.80\ 10^{-35}$&$1.32\ 10^{-35} $&$10^{11}$&$2.57\ 10^{-32}$&$1.09\  10
^{-32}    $ \\
      $10^5$&$1.91\ 10^{-34}$&$7.03\ 10^{-35}    $&$10^{12}$&$3.92\ 10^{-32}$&$1.67\ 10^{-32}$\\
      $10^6$&$6.87\ 10^{-34}$ &$2.65\ 10^{-34}  $&$10^{13}$&$ 5.68\ 10^{-32}$&$2.44\ 10^{-32} $\\
	 $10^{7}$&$1.94\ 10^{-33}$&$7.74\ 10^{-34}$&$10^{14}$&$7.92\ 10^{-32}$&$3.40\ 10^{-32}$\\ 
\hline
\end{tabular}

\end{center}
\end{table}
\def\arraystretch{1}  
{\em Robustness of cross sections:}  
The differential cross sections were evaluated numerically in Mathematica and found to be numerically stable, 
essentially independent of $Q^2_{\rm min}$ and the methods of integration.  The dependence of the cross sections 
on the functional form of $F_2^p(Q^2,x)$ for $1\ge x\ge x_P$ was tested by setting $F_2^p(Q^2,x)\sim x(1-x)^3$ 
for large $x$, and the change was found to be  $\sim 2$\% at $E_\nu=10^8$  and $\sim 0$ at $E_\nu=10^{12}$ GeV. 
If we set $F_2^p(Q^2,x)=0$ for $1\ge x\ge x_P$, an extreme case, we find the changes to be 6\% at $E_\nu=10^8$ GeV 
and $\sim 0$ at $E_\nu=10^{12}$ GeV. We tested our equipartition hypothesis by changing the strengths of the heavy 
sea quark distributions such that
\ba
s_s(x,Q^2)&=&\bar s_s(x,Q^2)=0.96 \,U(x,Q^2)\nonumber\\
c_s(x,Q^2)&=&\bar c_s(x,Q^2)=0.80 \,U(x,Q^2),\label{CTEQU}
\ea
similar to the distributions used by CTEQ.  This change gives us cross sections that are $\sim 6$\% greater at 
$E_\nu=10^8$ GeV and $\sim 3$\% greater at $E_\nu=10^{12}$ GeV. These variations are negligible compared to the very 
large differences with respect to the cross sections of Gandhi {\em et al.} \cite{gandhi} at the highest neutrino 
energies. Our calculations are numerically stable with regard to our choice of $x_{\rm min}$  in the integration, and 
thus, insensitive to our choice of $Q_{\rm min}=0.01$ GeV$^2$

{\em Conclusions:} 
We compute ultra-high energy neutrino cross sections based on an extrapolation to very small Bjorken $x$ of the 
logarithmic Froissart dependence in $x$ shown previously to provide an excellent fit to the measured proton structure 
function $F_2^p(x,Q^2)$ over a broad range of the virtuality $Q^2$.  In order to devise expressions for the neutral 
current and the charged current cross sections, we first extract quark and antiquark contributions based on a 
simple equipartition wee parton picture valid for $x_{\rm max}\la 10^{-3}-10^{-4}$ or 
$E_\nu^{\rm min}\ga 3\times 10^6-3\times 10^7$ GeV.  However, it is gratifying to see in Fig. \ref{fig:CC} that 
we are in excellent agreement 
with calculations based on CTEQ4-DIS parton densities over the much larger energy range $10\le E_\nu \le 10^8$ GeV.
The two sets of expectations diverge for $E_\nu \greaterabout 10^8$ GeV, as may be expected since our 
proton structure functions agree with those from CTEQ only for $x$-values greater than $10^{-3}$\cite{bbt}.  The 
increasing differences for $x < 10^{-3}$ reflect the fundamental difference in the assumed functional forms for 
the $x$ dependence, in our case a form that is constrained to increase no more rapidly than $\ln^2(1/x)$, in 
contrast to the inverse power growth in the CTEQ case.   
For large neutrino energies---above $10^9$ GeV---where much smaller $x$ is sampled, our Froissart-bound-model neutrino 
cross sections are as much as a {\em decade smaller} than those based on a pQCD extrapolation, a consequence of the fact 
that our structure function $F_2^p(x, Q^2)$ is significantly smaller at small $x$.   The very small $x$ region is also 
the region where our wee parton picture is most robust.  

The region of very small $x$ is a region of growing  
interest theoretically.  It is a region in which non-perturbative physics is expected to set in~\cite{glr-capella-soyez} 
and in which linear DGLAP pQCD evolution is not expected to hold.  While we cannot claim that logarithmic dependence 
on $x$ will result from a first-principles solution to small $x$ dynamics, neither can we 
expect an inverse power form to survive.  The logarithmic form we use 
provides an excellent fit to data over the range 
in $x$ and $Q^2$ where it has been tested. Its extrapolation to energies relevant in UHE neutrino studies provides 
estimates for event rates that should be taken into serious consideration for the planning and data analysis of new 
experiments.      
  
{\em Acknowledgments:} We thank the Aspen Center for Physics for hospitality during the research for this manuscript.  
The U.~S.\ Department of Energy, Division of High Energy Physics, under Contract No.\ DE-AC02-06CH11357 supports E.L.B.
The U.~S.~Department of Energy under Contract DE-FG02-91ER40688, TASK A supports in part C-I.T.  D.W.M. is supported 
by DOE Grant No. DE-FG02-04ER41308.  M.M.B thanks Prof. A. Vainstein of the University of Minnesota and Prof. L. Durand 
of the University of Wisconsin for both lengthy and fruitful discussions and encouragement.

\end{document}